\documentclass[10pt, conference, compsocconf]{IEEEtran}

\usepackage[pdftex]{graphicx}
\usepackage{bibspacing}
\usepackage{cite}
\usepackage{xcolor}
\usepackage{mdwmath}
\usepackage{mdwtab}
\usepackage[caption=false,font=footnotesize]{subfig}

\makeatletter
\newcommand*\titleheader[1]{\gdef\@titleheader{#1}}
\AtBeginDocument{%
	\let\st@red@title\@title
	\def\@title{%
		\bgroup\normalfont\large\centering\@titleheader\par\egroup
		\vskip1.5em\st@red@title}
}
\makeatother

\title{Video-Aware Measurement-Based Admission Control}
\titleheader{\textcolor{blue}{Published in:  2013 Australasian Telecommunication Networks and Applications Conference (ATNAC). Date Added to IEEE Xplore:  09 January 2014. Date of Conference: 20-22 Nov. 2013\\
		\textcopyright 2014 IEEE.  Personal use of this material is permitted.  Permission from IEEE must be obtained for all other uses, in any current or future media, including reprinting/republishing this material for advertising or promotional purposes, creating new collective works, for resale or redistribution to servers or lists, or reuse of any copyrighted component of this work in other works.\\
		DOI: 10.1109/ATNAC.2013.6705377}}

\author{\IEEEauthorblockN{Safeen Qadir}
	\IEEEauthorblockA{Electrical, Electronic and Computer Engineering\\
		University of Southern Queensland\\
		Queensland, Australia\\
		safeen.qadir@ieee.org}
	\and
	\IEEEauthorblockN{Alexander A. Kist}
	\IEEEauthorblockA{Electrical, Electronic and Computer Engineering\\
		University of Southern Queensland\\
		Queensland, Australia\\
		kist@ieee.org}}

\begin{document}

\maketitle

\thispagestyle{plain}
\pagestyle{plain}

%
\title{Video-Aware Measurement-Based Admission Control}


\maketitle

\begin{abstract}
Using instantaneous aggregate arrival rate as an admission control parameter will contribute to either bandwidth under-utilization or over-utilization. Being bursty in nature and variable in rate, video flows might encode any rate between a range of minimum and maximum values. At the time the decision is made, if the measured rate is at the minimum value, the bandwidth might be over-utilized due to accepting more sessions than the link can accommodate. In contrast, it might be under-utilized if the measured rate is at the maximum value due to rejecting more sessions than the link can accommodate.  The burstiness can be taken into account by considering the past history of the traffic. This paper investigates the suitability of the average aggregate arrival rate instead of the instantaneous aggregate arrival rate for video admission decisions. It establishes a mathematical model to predict the relationship between the two rates. Simulation results confirm that the average aggregate arrival rate is a more efficient decision factor for a small number of flows. Although it has no additional advantage for moderate and large number of flows, it still can stabilize the admission decision by smoothing the burstiness of a set of the instantaneous rates (within the measurement period) over a period of time.


\end{abstract}

\begin{IEEEkeywords}
Admission Control; Measurement-Based Admission Control; Video;

\end{IEEEkeywords}

\IEEEpeerreviewmaketitle

\section{Introduction}
In the foreseeable future, realtime video traffic will become the dominant network traffic. Cisco predicts that "It would take over 6 million years to watch the amount of video that will cross global IP networks each month in 2016. Every second, 1.2 million minutes of video content will cross the network in 2016" \cite{Cisco2012}. The actual figure of real-time entertainment services such as Hulu and Netflix in north America in 2011 was 58.6\% of the total Internet traffic \cite{IEEESpectrum_Jan2013}.

To avoid congestion for non-adaptive traffic, binary-based admission control is the dominant technique \cite{Latre2011}. It either allows or blocks new traffics based on available resources. Inelastic traffic specifies the maximum and minimum bitrates during the admission process. The network nodes police the maximum rate to ensue that it is not exceeded, and they attempt to guarantee the transmission of the minimum bit rate. This kind of admission controls (which reserve a fixed amount of resources for each session) is suitable for services such as voice telephony that have a constant-bit rate. However, due to the fact that each video sequence is encoded differently based on the complexity of the video image, these are ineffective for video traffic with bursty bit rates. Furthermore, rate adaptation which contributes potentially to the increase of the burstiness and rate variation has been proposed to optimize video quality and the Quality of Experience (QoE) \cite{Rengaraju2012,Khalek2012,Politis2012,Qadir2013}.

Since video traffic is sent at variable rate, the aggregate rate is considerably lower than the sum of peak rates \cite{Nevin2010} and therefore admission decisions shouldn't be based on worse-case bounds. This will achieve the optimum trade-off between utilization and perceived QoE. 

To protect the quality of video over Internet, there is a necessity for an admission control at the edge of network. Furthermore, to take advantage of the bursty nature of video traffic and to accept more sessions, the acceptance/denial decision of the admission control can be based on the history of video rate instead of the instantaneous rate. This is due to the fact that the burstiness of video flows can be compensated by the silence period of other flows. Considering past video rate will give a better indication of the behaviour of video flows and potentially is a better reference rate for the admission control to rely on.

The primary aim of measurement-based admission control (MBAC) is to eliminate or reduce the need of flow state information and control overhead for admission decision, and to maximize utilization at an eventual cost of QoS degradation \cite{Lima2007}. Few MBAC procedures have been proposed \cite{Breslau2000}, however there are still a few which simply rely on the instantaneous aggregate arrival rate.  Pre-Congestion Notification (PCN)-based admission control which recently has been standardized by the IETF depends on the instantaneous admissible and supportable rates for its operation \cite{EardleyP.2009}. Despite all the efforts, there is still no entirely satisfactory admission algorithm for variable rate flows \cite{Auge2011}. 

This paper investigates the suitability of the average aggregate arrival rate instead of the instantaneous aggregate arrival rate for video streaming admission decision. Furthermore, the probability relationship between both rates is established mathematically. The rest of the paper is organized as follows. Assumptions used in this study are justified in Section II, and related work is reviewed in Section III. Section IV presents the mathematical model of the proposed scheme, and the simulation setup is explained in Section V. The results are presented and discussed in Section VI. The paper is concluded in Section VII.


\section{Assumptions}

\noindent This papers makes the following assumptions:
\begin{itemize}
\item Video traffic is the dominant Internet traffic \cite{Cisco2012}. Other traffic forms small percentage and therefore only video traffic will be considered.
\item To provide required level of QoE, explicit admission control is required \cite{Nevin2010}.
\item This paper contributes to the measurement mechanism of admission control procedure, hence metering and marking of packets are not discussed.
\end{itemize}

\section{Related Work}
In this section, we review the MBACs that depend on per-aggregate rather than per-flow. We refer interested readers of recent admission control procedures and classifications to \cite{Menth2010} \cite{Lima2007} \cite{Wright2007}. The Measured Sum (MS) algorithm was proposed in \cite{Jamin1997} which uses measurement to estimate the load of existing traffic. A new flow will be admitted if the sum of the required rate of the new flow and estimated rate of existing flows is less than a utilization target times the link bandwidth. Four measurement-based admission control algorithms are presented in \cite{Gibbens1997} based on Chernoff bounds.

\cite{Floyd1996} proposed an admission control scheme for controlled-load services that estimates the equivalent capacity of a class of aggregated traffic based on Hoeffding bounds. The paper concluded that equivalent capacity based admission is efficient for classes with as few as 50 connections however is the same as a peak-rate admissions control procedure for classes with only 10 connections. The paper also presented a formulation of equivalent capacity that is suitable for classes with either a moderate number of admitted connections or a wide range in the peak rates of the admitted connections. 

An extension to the PCN-based admission control system has been proposed in \cite{Latre2011a}. A novel metering algorithm based on a sliding-window to cope with the bursty nature of video sessions and another adaptive algorithm to facilitate the configuration of the PCN were proposed.

\cite{Nevin2010} studied how uncertainty in the measurements of MBAC vary with the length of the observation window and described a methodology for analyzing measurement errors and performance. The concept of similar flows and adding a slack in bandwidth were introduced to minimize the probability of false acceptance. 

\cite{Auge2011} proposed a MBAC scheme based on measured mean and variance of load offered to the cross-protect priority queue. \cite{Moore2002} performed an implementation-based comparison of MBAC algorithms using a purpose built test environment. The study revealed that there is no single ideal MBAC algorithm due to computation overheads, multiple timescales present in both traffic and management, and error resulting from random properties of measurements which dramatically impact the MBAC algorithm's performance.

\section{Modeling}
This section presents a mathematical model for the proposed average aggregate rate-based admission control. We consider a network of \emph{N} nodes and \emph{M} $\subseteq$ \emph{N} $\times$ \emph{N} links, where link \emph{l} $\in$ \emph{M} and \emph{F} denotes the set of flows where \emph{f} $\in$ \emph{F}.



Assuming that requests for video sessions are independent random variables, the instantaneous aggregate arrival rate \emph{$X_{inst}(l,t)$} of all flows \emph{F} on link \emph{l} at time \emph{t} is

\begin{equation}
\label{eq:1}
X_{inst}(l,t) = \sum_{i=1}^n x_{i}(l,t)
\end{equation}
for \emph{i} $>$ 0 and \emph{t} $>$ 0. Where \emph{$x_i(l,t)$} is the instantaneous arrival rate of session \emph{i} on link \emph{l} at time \emph{t}, and \emph{n} is the number of sessions.

In current admission control approaches, a new session will be accepted if the sum of arrival rates \emph{f} plus the peak rate of the new session \emph{$x_{new}$} is less or equal to the link's capacity \emph{$C_{l}$} as given by Equation (\ref{eq:2}).

\begin{equation}
\label{eq:2}\
X_{inst}(l,t) + x_{new} \le  \emph{C}_{\emph{l}}
\end{equation}

The proposed scheme considers the average aggregate arrival rate of active sessions as an admission parameter instead of the instantaneous aggregate arrival rate.

Now we find how the instantaneous arrival rate is related to it's average. Let 
\emph{$x_i(l,t)$} be an independent random variable with minimum rate \emph{$x_i^{min}(l,t)$}, peak rate \emph{$x_i^{max}(l,t)$}, and \emph{$x_i^{min}(l,t)$} $\le$ \emph{$x_i(l,t)$} $\le$ \emph{$x_i^{max}(l,t)$}.

The Hoeffding inequality theorem \cite{Hoeffding1963}, the probability that 
\emph{$X_{inst}(l,t)$} exceeds its mean $\mu_r(l,\tau)$ by a positive number \emph{n}$\epsilon$  for $\epsilon$ $>$ \emph{0}, is given by Equation (\ref{eq:3}):

\begin{equation}
\label{eq:3}
Pr\{X_{inst}(l,t) \ge \mu_{r}(l,\tau)+n\epsilon\} \le \delta
\end{equation}

Where $\delta$ is given by Equation (\ref{eq:4})

\begin{equation}   
\label{eq:4}\
\delta = exp \left(   \frac{-2n^2\epsilon^2} { \sum_{i=1}^n (x_{i}^{max}(l,t)-x_{i}^{min}(l,t))^2} \right)  
\end{equation}

and $\mu_r(\emph{l},\tau)$ is the average aggregate arrival rate which is the amount of data that arrives in the interval [\emph{$t_1,t_2$}]. It can be calculated for the period $\tau= {t_2}$ - {$t_1$} (\emph{$t_2 > t_1$}) when a new session is requested at time \emph{$t_2$} using Equation (\ref{eq:5}). 

\begin{equation}   
\label{eq:5}\
\mu_r(l,\tau) = \frac{1}{\tau}  \int_{t_1}^{t_2} \sum_{i=1}^n x_{i}(l,t) \  dt
\end{equation}


Hence, a new requested session will be accepted if the condition in Equation (\ref{eq:6}) is met:

\begin{equation}
\label{eq:6}
\mu_r(l,\tau) + x_{new} \le  C_{l}
\end{equation}

\noindent Substituting Equation (\ref{eq:5}) in Equation (\ref{eq:6}) yields:
\begin{equation}
\label{eq:7}\
\frac{1}{\tau}  \int_{t_1}^{t_2} \sum_{i=1}^n x_{i}(l,t) dt + x_{new} \le  C_{l}
\end{equation}

where \emph{x}$_{new}$ is the requested rate of the new session, and \emph{C}$_{\emph{l}}$ is the link bandwidth. Unlike the \emph{x}$_{new}$ in Equation (\ref{eq:2}), we propose to be set based on the required QoE. Setting a proper value of \emph{x}$_{new}$ ensures QoE experienced by the end user. We propose  per video quality class for \emph{x}$_{new}$ such as 11Mbps for full HD, 8Mbps for HD ready, 2Mbps for SD and 1.25 for HD web. This approach does not require keeping per-flow peak rate for each individual flow as proposed in \cite{Floyd1996} which causes scalability issues.



As the Hoeffding theorem limits the upper bounds of the probability that $\mu_r(\emph{l},\tau)$ is less than \emph{$X_{inst}(l,t)$}, the proposed scheme in Equation (\ref{eq:6}) makes use of $\mu_r(\emph{l},\tau)$ to ensure the maximum possible number of sessions accepted utilizing the bandwidth efficiently comparing to the admission procedure in Equation (\ref{eq:2}) which relies on the instantaneous aggregate rate \emph{$X_{inst}(l,t)$}. Setting the proper value of \emph{x}$_{new}$ ensures that the required QoE is provided.

\section{Simulation Setup}

Matlab was used to evaluate the proposed scheme. Five different video trace libraries (Tokyo Olympics, Silence of the Lambs, Star Wars IV, Sony Demo and NBC News) from publicly available libraries \cite{Seeling2004} \cite{VanderAuwera2008} were used. The aim was to have different video contents such as slow moving pictures (news) and fast moving pictures (sports).

The frames were randomly picked from the libraries and the rates were measured for a period of five frames. The instantaneous aggregate rate was measured at the end of each period and the average aggregate rate was measured over the period. As the instantaneous rate varies, it may have any value within the measurement period. 



In order to see how the number of flows influences the probability relationship between the instantaneous and average aggregate arrival rates, different number of flows were simulated. For each flow, the measurements were taken five times, 100 random runs of the scheme for each time. Then the probability that the average rate is less than the instantaneous rate was calculated for each run. 

\section{Results and Discussion}

\begin{figure}[!t]
\centering
\includegraphics[width=\linewidth]{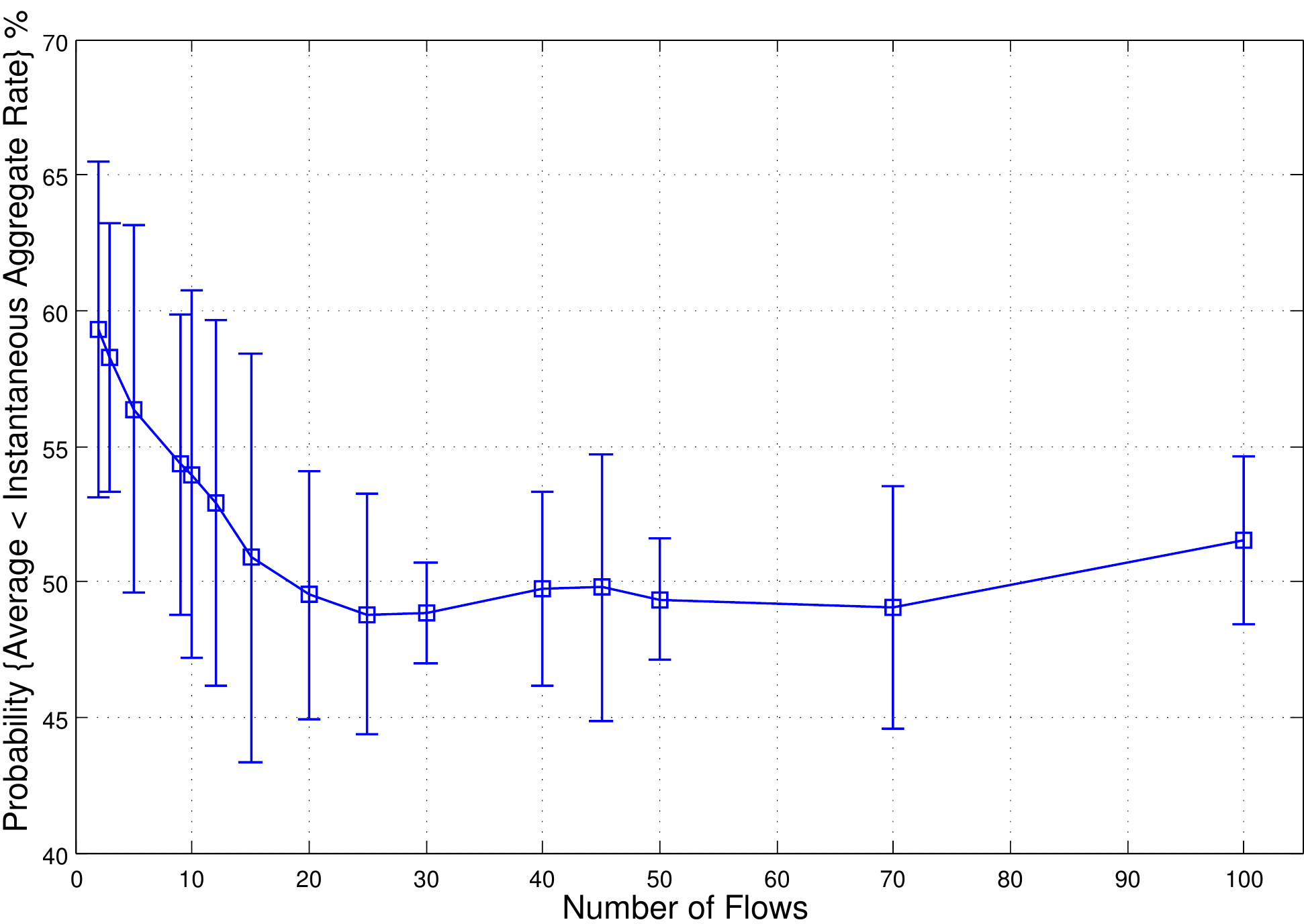}
\caption{Average and Instantaneous Aggregate Arrival Rates Relationship}
\label{probabilityRelationship}
\end{figure}

The mean and the confidence interval of the probability relationship between the average and instantaneous aggregate rates are plotted against the number of flows in Figure \ref{probabilityRelationship}. The overall trend of the curve is decreasing which means that the probability that the average aggregate arrival rate is less than the instantaneous aggregate arrival rate decreases with the increase of the number of flows. For as few as 15 flows, the average has advantage over the instantaneous rate, accepting more flows. The exponential shape-like of the probability in Figure \ref{probabilityRelationship} can be validated by the exponential relationship in Equation (\ref{eq:4}).

As the number of flows increases, both rates approaching each other which indicates that there is no difference in considering either rate. The probability fluctuates around 50\% for above 15 flows which produces uncertainty in the instantaneous rate. However, considering the average rate for any number of flows will still contribute in decreasing the burstiness of a set of instantaneous rates within the measurement period and adding more consistency in admission decision.

\begin{figure}[!t]
\centering
\includegraphics[width=\linewidth]{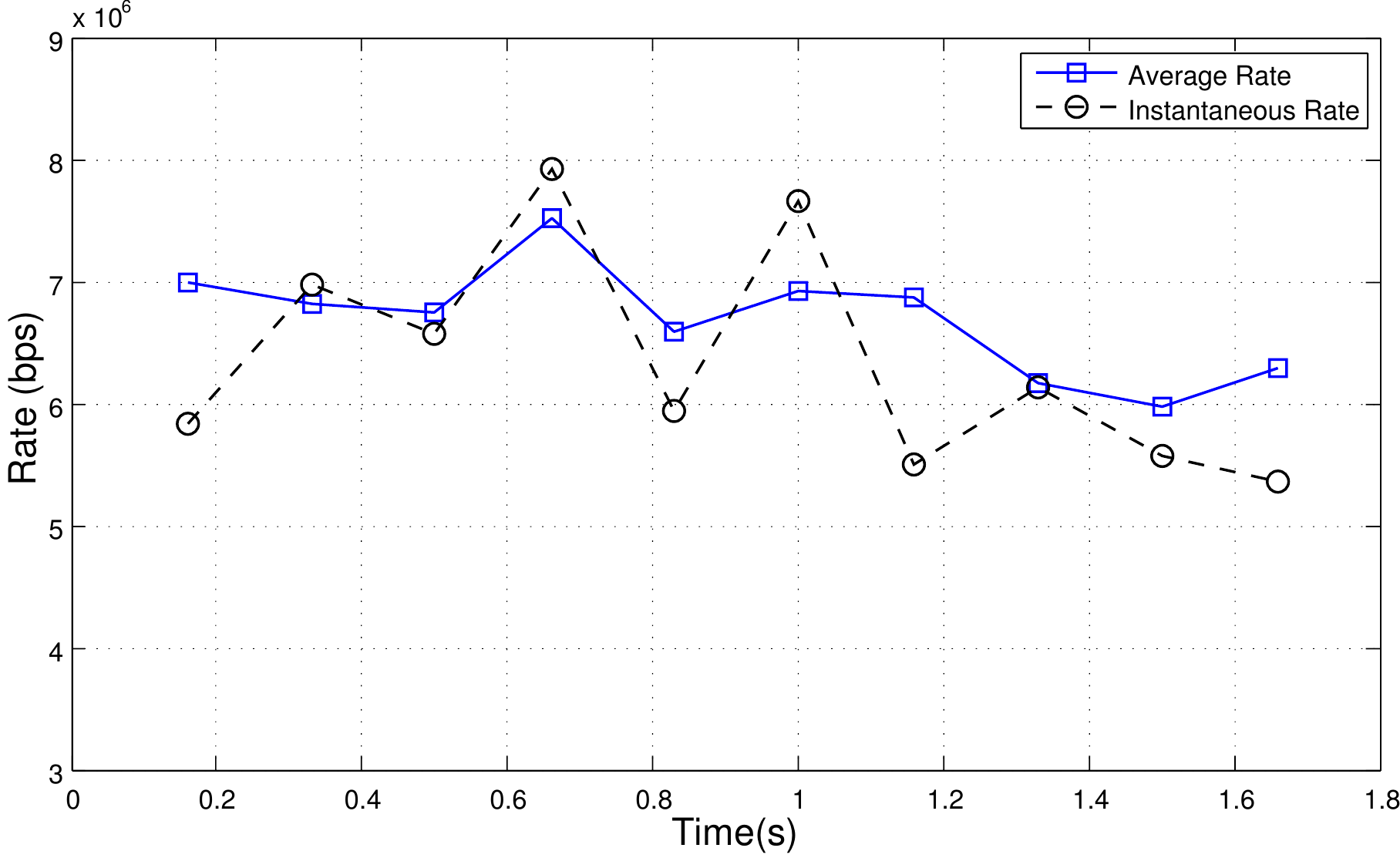}
\caption{Average and instantaneous aggregate rates - 5 flows}
\label{Fluctuation_5_New}
\end{figure}

\begin{figure}[!t]
\centering
\includegraphics[width=\linewidth]{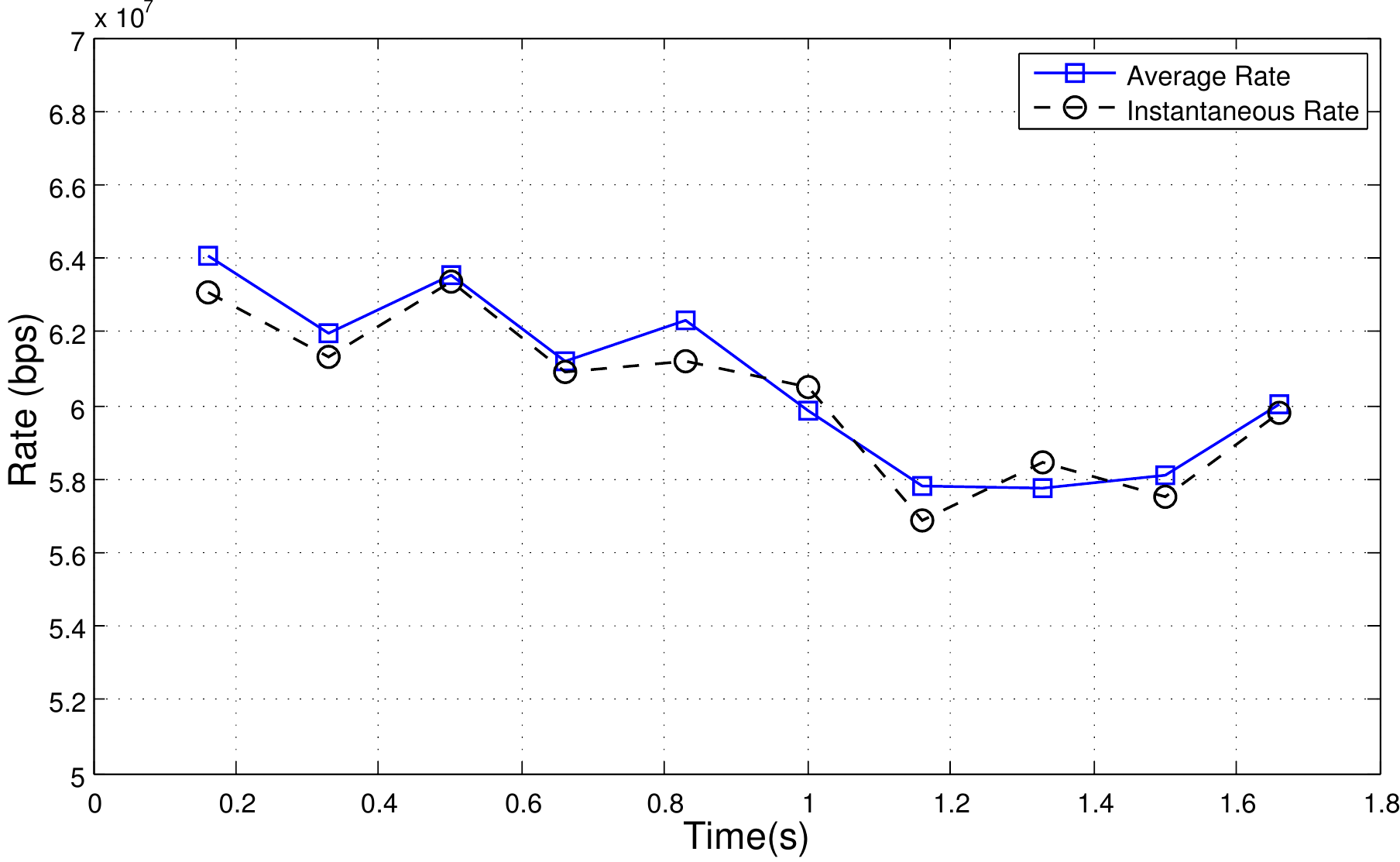}
\caption{Average and instantaneous aggregate rates - 40 flows}
\label{Fluctuation_40_New}
\end{figure}

Figures \ref{Fluctuation_5_New} and \ref{Fluctuation_40_New} show the average and instantaneous aggregate rates for 5 and 40 flows. The smoothness is more observable in the case of 5 flows because the overall rate gets smoother (less bursty) in the case of 40 flows. Table \ref{table:burstiness} compares the burstiness of both rates for each of 5 and 40 flows using the peak-to-mean ratio and coefficient of variance methods \cite{Frost1994}. As shown in the table, the burstiness of the average is less than the instantaneous and decreases with the increase of the number of flows. It also can be noticed that the burstiness of both rates reach the same value for large number of flows (40 flows in this paper).

\begin{table}[!th]
\centering
\caption{The Burstiness of the Average and Instantaneous Aggregate Rates}
\label{table:burstiness}
\begin{tabular}{|c|c|c|c|c|}
                                                                         \hline
   & \multicolumn{2}{|c|}{{Peak-to-Mean Ratio}}  &    	\multicolumn{2}{|c|}{{Coefficient of Variation}} \\\hline   
Rate & 5 flow & 40 flow & 5 flow & 40 flow\\\hline                                           
Average & 1.12 & 1.05 & 0.3 & 0.03 \\\hline
Instantaneous & 1.24 & 1.05 & 0.33 & 0.03 \\\hline
\end{tabular}
\end{table}

\begin{figure}[!t]
\centering
\includegraphics[width=\linewidth]{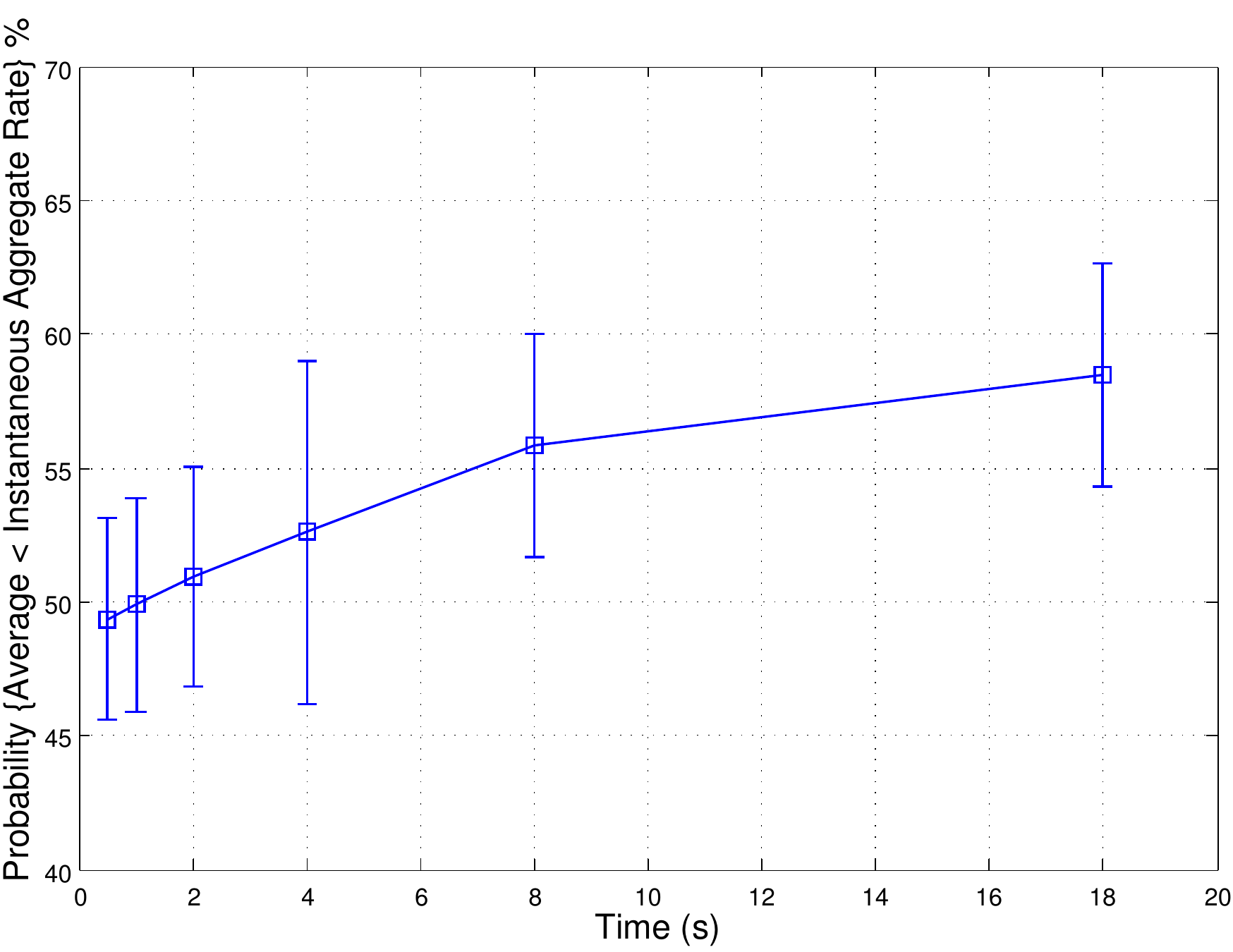}
\caption{The increase of probability for longer measurement time period - 40 flows}
\label{timescale_40}
\end{figure}

Further simulations were performed to investigate the impact of the measurement period on the probability. Figure \ref{timescale_40} shows the mean and confidence interval of the probability for different measurement time periods for 40 flows scenario. It can be seen that it is higher for longer period than the shorter. On the one hand, considering longer measurement period will provide higher probability that the average is less than the instantaneous rate resulting in more sessions to be accepted. On the other hand, it makes admission control to be less reactive to the changes of traffic rate.

\begin{figure}[!t]
\centering
\includegraphics[width=\linewidth]{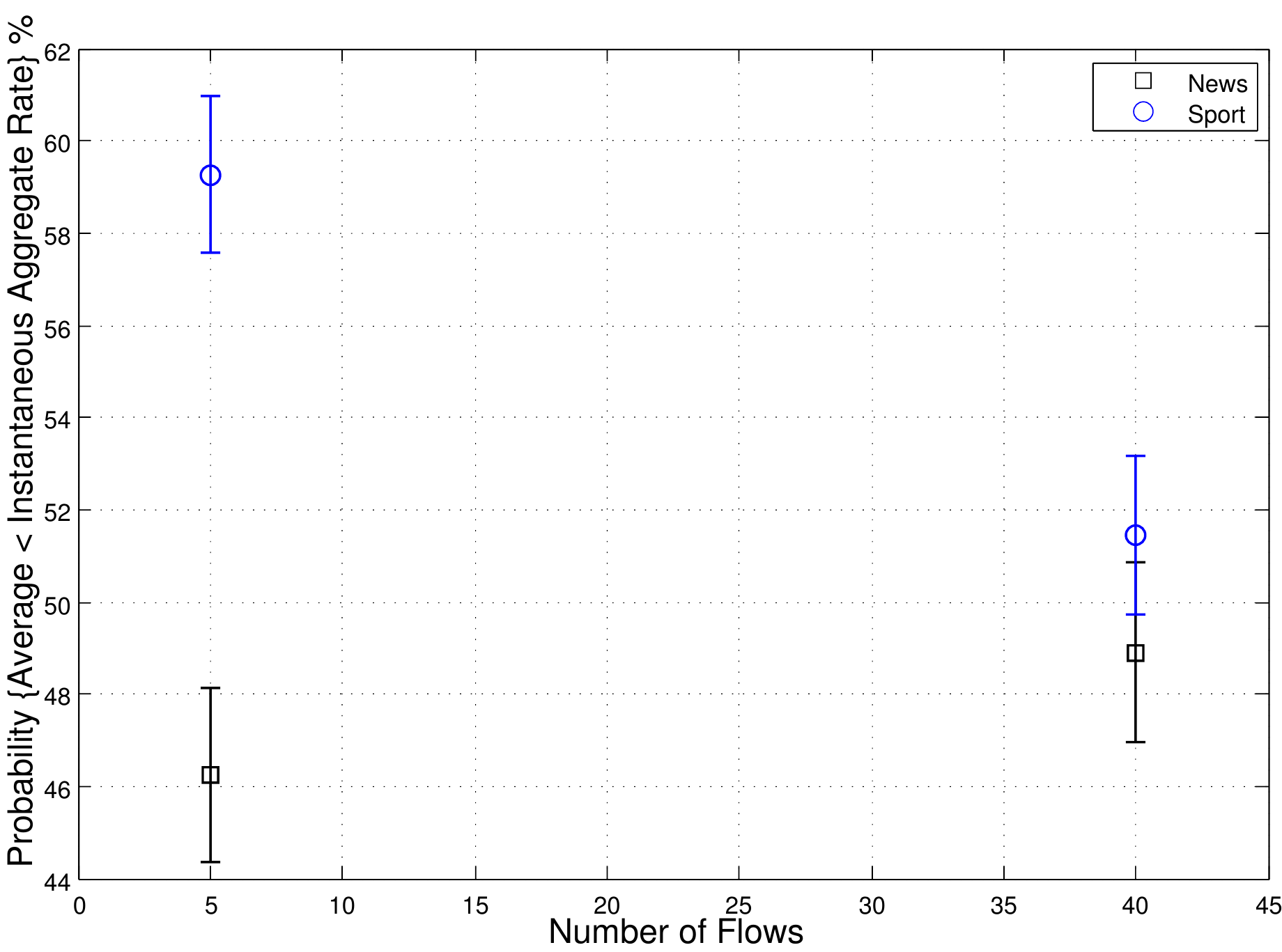}
\caption{Probability for news and sports - 5 and 40 flows}
\label{probability_content}
\end{figure}
The video content also has impact on the probability relationship between the two rates. Figure \ref{probability_content} shows a significant difference between news (46\%) and sports (\%59) for 5 flows, while a small change from 49\% (news) to 51\% (sport) for as big as 40 flows. This indicates the suitability of using the average rate for small number of fast moving video scenes such as sports than slow moving video scenes such as news. 


The uncertainty in the instantaneous arrival rate which is essentially caused by the burstiness of video traffic and/or rate adaptation strategy, will contribute negatively to decisions made by admission control procedures. At the time of decision, if the measured rate is at the minimum value, the bandwidth might be over-utilized due to accepting more sessions than the link can accommodate. In contrast, it might be under-utilized if the measured rate is at the maximum value due to rejecting more sessions than the link can accommodate. 

To avoid this scenario and utilize bandwidth more efficiently, the average aggregate arrival rate over a period of time is a more efficient decision factor to be taken for a small number of flows. Thus more flows are admitted and bandwidth is utilized more efficiently. This is due to the fact that the burstiness of the average aggregate rate can be compensated by the silence period over time.

The rate uncertainty mentioned earlier, under and over-utilization of bandwidth can be controlled by considering the past history of the traffic using the above proposed strategy. By this strategy, the considered rate for admission decision reflects a  better characteristic of the nature of video flows.

\section{Conclusion}
The suitability of using the average aggregate arrival rate instead of the instantaneous rate was investigated for video admission decision. A mathematical model for the probability relationship between both rate was formulated. The simulation results have shown higher probability that the average is smaller than the instantaneous aggregate arrival rate for as few as 15 flows. Whereas there is no significant difference between the two rates for higher number of flows, the average rate will be smoother than the individual instantaneous rate within the measurement period that stabilizes the admission decision. This work has found that the probability relationship is impacted by fast moving video scenes such as sports more than slow moving video scenes such as news. Real time simulation will be conducted in the future to implement the admission criteria in Equation (\ref{eq:7}) at the packet level. Furthermore, the performance of the proposed scheme will be evaluated based on the number of admitted session and perceived QoE of each session.


\end{document}